\pdfoutput=1

\documentclass[twocolumn]{revtex4}

\usepackage{ifthen}	
\newboolean{isnature}
\setboolean{isnature}{false}

\usepackage[latin1]{inputenc}
\usepackage{graphicx}
\usepackage{amsmath,amssymb}
\usepackage{hyperref}
\usepackage{latexsym}
\usepackage[squaren,thickspace,thickqspace]{SIunits}
\usepackage{multirow}
\usepackage{textcomp}
\usepackage{pdflscape}
\usepackage{rotating}
\usepackage{pbox}
\usepackage{ifthen}	

\newcommand{\nbar}{\bar{n}}

\newcommand{\caf}{\ensuremath{^{40}{\rm Ca}^{+}\, }}

\newcommand{\create}{\ensuremath{{\,\hat{a}^{\dagger}}}}
\newcommand{\destroy}{\ensuremath{\,\hat{a}}}

\newcommand{\splus}{\ensuremath{\hat{\sigma}_+}\,}

\newcommand{\D}{\mathrm{d}}
\newcommand{\I}{\mathrm{i}}
\newcommand{\Exp}[1]{\mathrm{e}^{#1}}
\newcommand{\der}[3][1]{\ifthenelse{#1=1}{\frac{\D#2}{\D#3}}
	{\frac{\D^#1#2}{\D#3^#1}}}%
\newcommand{\pder}[3][1]{\ifthenelse{#1=1}
	{\frac{\partial #2}{\partial #3}}{\frac{\partial ^#1#2}{\partial #3^#1}}}%
\renewcommand{\vec}[1]{{\boldsymbol{#1}}}
\newcommand{\op}[1]{\hat{#1}}
\newcommand{\bra}[1]{\bigl<#1\bigr|}
\newcommand{\ket}[1]{\bigl|#1\bigr>}
\newcommand{\rbra}[1]{\left(#1\right)}
\newcommand{\sbra}[1]{\left[#1\right]}
\newcommand{\cbra}[1]{\left\{#1\right\}}
\newcommand{\abs}[1]{\left| #1 \right|}
\let\FPi\Pi	\renewcommand{\Pi}{\mathit{\FPi}}
\newcommand{\be}{\begin{eqnarray}}
\newcommand{\ee}{\end{eqnarray}}

\begin{document}

\newcommand{\Title}{Fast quantum control and light-matter interactions at the 10,000 quanta level}

\newcommand{\capfigi}{
	\ifthenelse{\boolean{isnature}}{\begin{figure*}}{\begin{figure}}
		\includegraphics[width = 1\columnwidth]{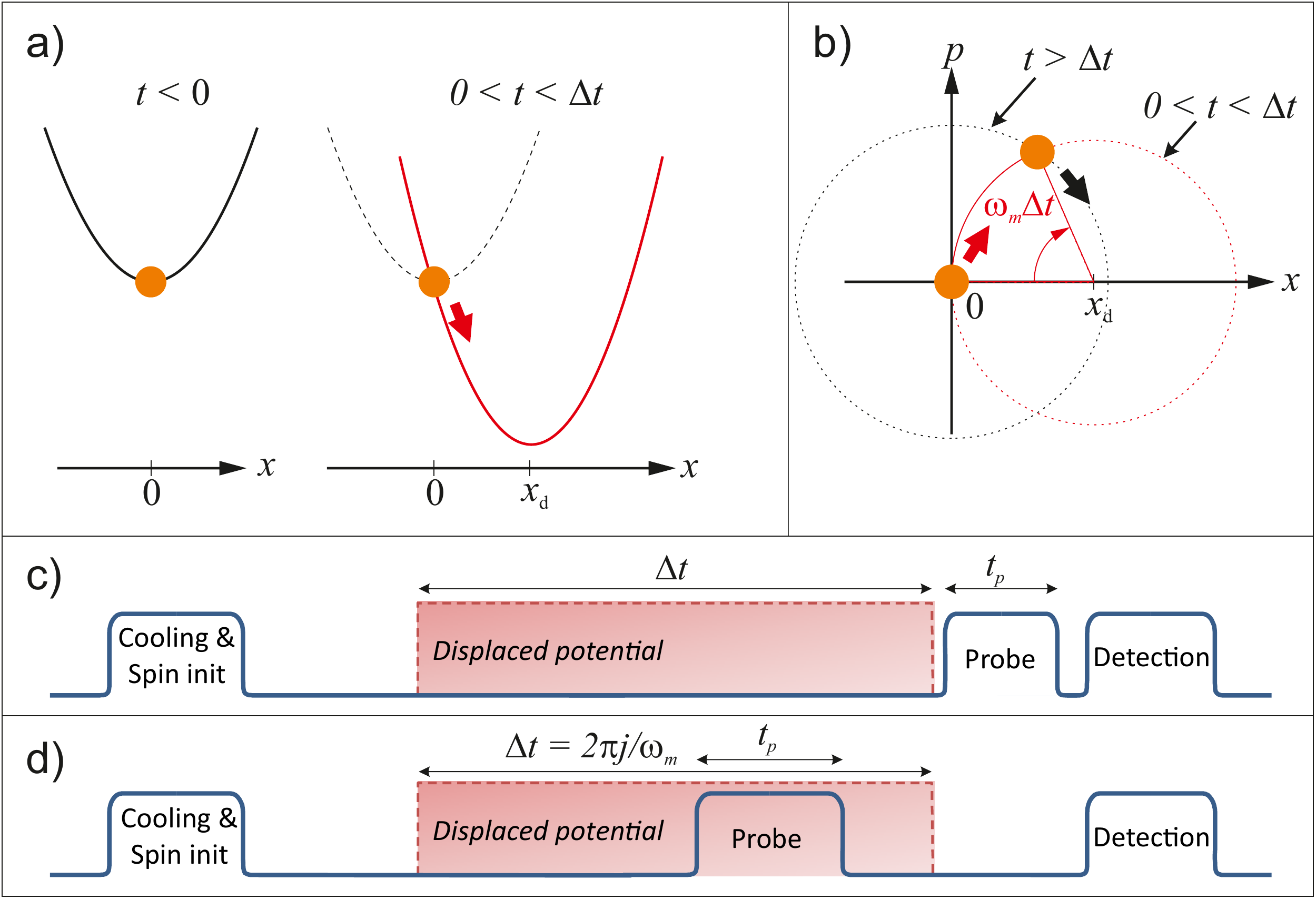}
		\caption{\textbf{Experimental sequences:} a) The ion is first cooled close to the ground state of motion in a harmonic trapping potential centered at $x=0$, and the spin state is pumped to $\ket{\downarrow}$ (see Supplementary Information). At $t=0$ the potential center is suddenly displaced to $x_\text{d}$ and the ion's motional state becomes $\ket{-\alpha_0}$. b) Evolution of the motional wave function in the position-momentum phase-space. After a time $\Delta t$ in the displaced potential, the trap center is switched back to $x=0$, leaving the ion in a coherent state of size given by equation (\ref{eq:alpha}). c) Sequence of laser-pulses and potential displacements for monitoring the time evolution of coherent states. d) Sequence for studying the light-matter interaction with highly-excited states. In this sequence, $\Delta t$ is an integer number of times $j$ of the oscillator period.
		}
		\label{fig:expseq}
	\ifthenelse{\boolean{isnature}}{\end{figure*}}{\end{figure}}
}

\newcommand{\capfigii}{	
	\begin{figure*}
		\includegraphics[width = 1.95\columnwidth]{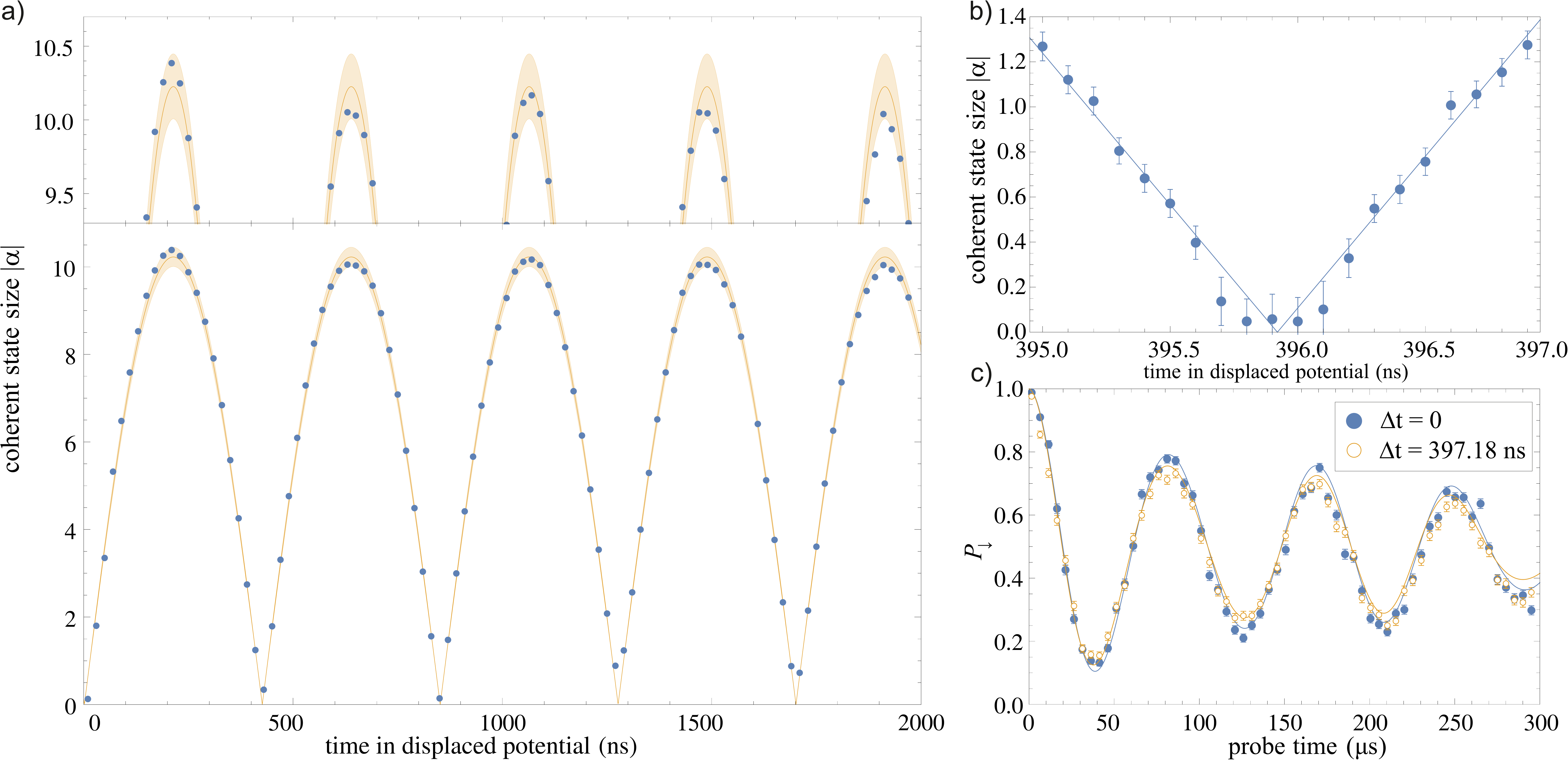}
		\caption{\textbf{Coherent-state generation:} a) The states are prepared in the original potential by executing the experimental sequence in Fig.\ \ref{fig:expseq}c. The solid orange line is a fit to the data using equation (\ref{eq:alpha}). The fluctuations in amplitude can be explained by drifts in the laser power over the time taken to acquire the data (shaded area, see Supplementary Information). b) Sub-nano-second resolution scan around one complete motional period ($\Delta t\approx2\pi/\omega_\text{m}$). This measurement was performed with $\omega_\text{m}/(2\pi)\approx\unit{2.53}{MHz}$ and corresponds to $\alpha_0\approx 85$. c) Measured probability of finding the spin in the $\ket{\downarrow}$ state as a function of the probe time using the $s=1$ sideband. Filled symbols correspond to the initially-prepared state and open symbols after approximately one motional period. This measurement was performed with $\omega_\text{m}/(2\pi)\approx\unit{2.52}{MHz}$ and $\nbar_\text{th}\approx 0.15$, and corresponds to $\alpha_0\approx 85$. Every data point consists of 1000 measurements. Error bars are assigned assuming quantum projection noise according to Ref. \cite{95Itano}.}
		\label{fig:lobes}
	\end{figure*}
}	

\newcommand{\capfigiii}{	
	\begin{figure*}
		\includegraphics[width = 1.95\columnwidth]{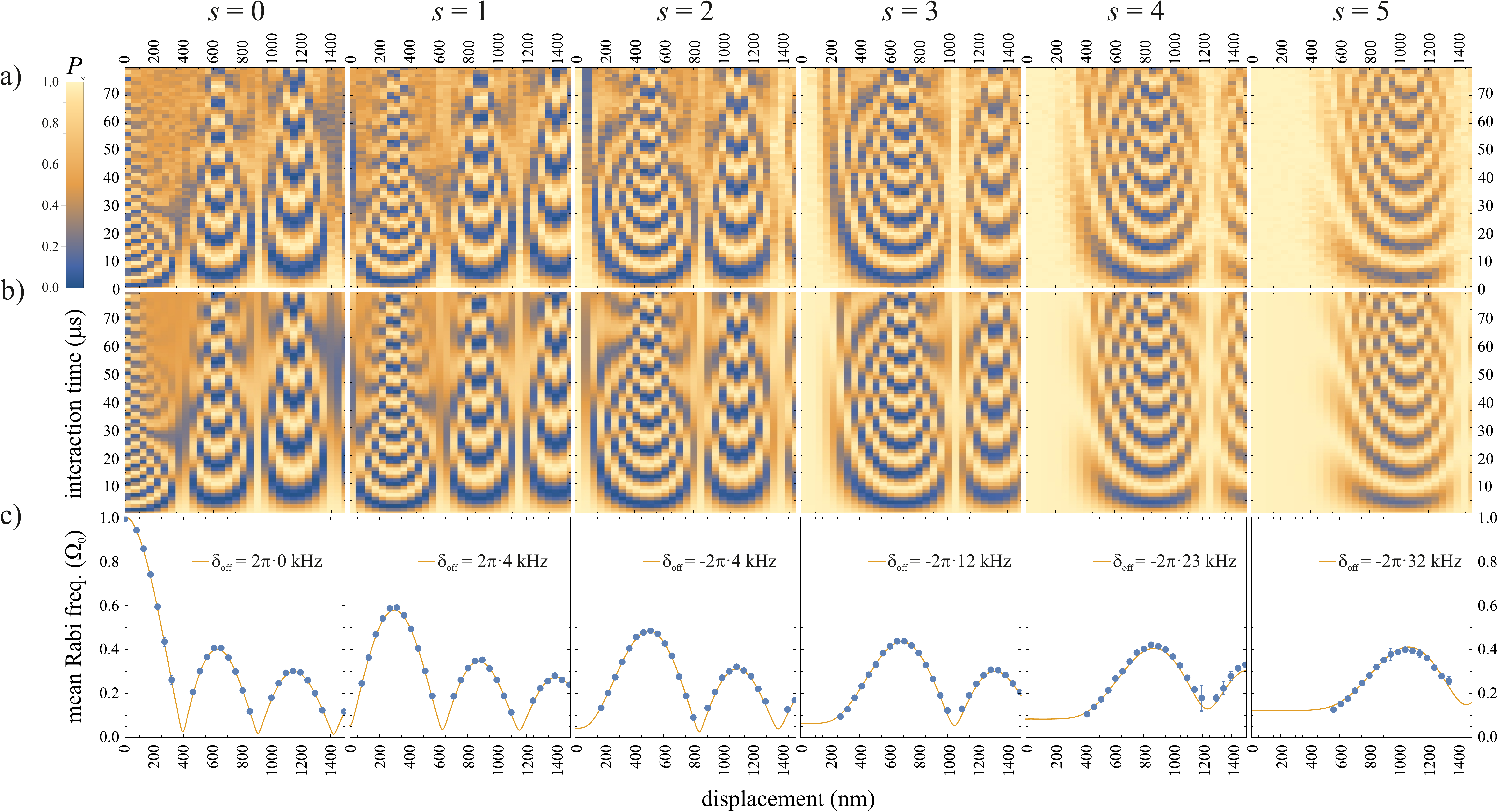}
		\caption{\textbf{Light-atom interaction:} a) Measurement of the probability $P_\downarrow$ of finding the spin in the $\ket{\downarrow}$ state as a function of the interaction time with light near-resonant with the $s^\text{th}$ motional sideband and of the trap-potential displacement $x_\text{d}$. Our computer-control system imposes $t_\text{p}\geq\unit{1.4}{\micro s}$, so for high Rabi frequencies $P_\downarrow$ can be close to zero even for the first point. b) $P_\downarrow$ obtained using equation (\ref{eq:Pdispth}) with the parameters determined from fits. c) Mean Rabi frequencies for displaced thermal states as a function of the potential displacement. The data points are given by the center of the Lorentzian curves to which we fit $\tilde{P}_\downarrow(\Omega)$, the Fourier transform of $P_\downarrow(t_\text{p})$. Most error bars are hidden behind the points. The solid lines result from the theory model including AC Stark shifts and a detuning $\delta_\text{off}$ obtained from fitting the data for a given $s$.}
		\label{fig:2Dplots}
	\end{figure*}
}

\ifthenelse{\boolean{isnature}}
{\title{\Title}
	\author{J. Alonso$^{1}$\footnote{These authors contributed equally to this work}, F. M. Leupold$^{1*}$, Z. U. Sol\`er$^1$, M. Fadel$^1$\footnote{Present address: Department of Physics, University of Basel, Klingelbergstrasse 82, 4056 Basel, Switzerland}, M. Marinelli$^1$, B. C. Keitch$^1$\footnote{Present address: Department of Engineering Science, University of Oxford, Parks Road, Oxford OX1 3PJ, UK}, V. Negnevitsky$^1$, J. P. Home$^1$}
	\maketitle
	\begin{affiliations}
		\item Institute for Quantum Electronics, ETH Z\"urich, Otto-Stern-Weg 1, 8093 Z\"urich, Switzerland
	\end{affiliations}
}
{\title{\Title}
	\author{J. Alonso\footnote{These authors contributed equally to this work}}\email{alonso@phys.ethz.ch}
	\author{F. M. Leupold$^\ast$}
	\author{Z. U. Sol\`er}
	\author{M. Fadel\footnote{Present address: Department of Physics, University of Basel, Klingelbergstrasse 82, 4056 Basel, Switzerland}}
	\author{M. Marinelli}
	\author{B. C. Keitch\footnote{Present address: Department of Engineering Science, University of Oxford, Parks Road, Oxford OX1 3PJ, UK}}
	\author{V. Negnevitsky}
	\author{J. P. Home}\email{jhome@phys.ethz.ch}
	\affiliation{Institute for Quantum Electronics, ETH Z\"urich, Otto-Stern-Weg 1, 8093 Z\"urich, Switzerland}
	\maketitle
}

\textbf{Fast control of quantum systems is essential in order to make use of quantum properties before they are degraded by decoherence. This is important for quantum-enhanced information processing \cite{BkNielsen}, as well as for pushing quantum systems into macroscopic regimes at the boundary between quantum and classical physics \cite{03Zurek,07Zurek}. ``Bang-bang'' control attains the ultimate speed limit by making large changes to  control fields on timescales much faster than the system can respond \cite{99Viola,99Vitali},  however these methods are often challenging to implement experimentally. Here we demonstrate bang-bang control of a trapped-ion oscillator using nano-second switching of the trapping potentials. We perform controlled displacements which allow us to realize quantum states with up to 10,000 quanta of energy. We use these displaced states to verify the form of the ion-light interaction at high excitations which are far outside the usual regime of operation. These methods provide new possibilities for quantum-state manipulation and generation \cite{09Serafini,09Serafini2}, alongside the potential for a significant increase in operational clock speed for ion-trap quantum information processing \cite{13Alonso, 98Wineland2}.}

Quantum control involves the preparation, manipulation and exploitation of high-purity quantum states, which constitute a fundamental resource for quantum information processing \cite{BkNielsen}. For quantum harmonic oscillators, experiments have demonstrated such control using confined optical and microwave fields \cite{96Brune,99Parkins}, and using the mechanical oscillations of single trapped ions \cite{96Meekhof, 10Zahringer,15Kienzler,15Lo}. Most of this work was performed in the adiabatic and resonant regimes, where control timescales are much slower than the system's natural frequencies. ``Bang-bang'' control is the opposite extreme, in which timescales are so short that changes to the system can be considered to be instantaneous \cite{99Viola,06Morton,09Damodarakurup}. For trapped ions, bang-bang control has been performed using ultra-fast laser pulses \cite{10Campbell,13Mizrahi}. However, optical fields interact with the electric dipole or quadrupole moment of the atom, and the oscillator states prepared in this way have thus far been limited to excitations of less than 20 quanta. Electric fields acting directly on the charge (electric monopole) can be used for achieving a strong-interaction regime. This is a key element for transporting trapped-ion qubits in a scalable quantum-computing architecture \cite{98Wineland2, 02Kielpinski}, where operations have been performed on timescales comparable to the oscillator frequency \cite{12Bowler,12Walther}. In these examples the speed was limited by the bandwidth of the filtered control voltages applied to the electrodes.

\ifthenelse{\boolean{isnature}}{}{\capfigi}

In this Letter, we demonstrate bang-bang control over trapped-ion harmonic oscillators, which we use to generate and characterize quantum states with up to 10,000 quanta. To achieve this, we wire the trap electrodes according to a new scheme: we prepare constant, filtered voltages at the inputs of switches placed in vacuum close to the trap electrodes, and select the input to be wired to the electrode by means of digital control pulses (for further details about the experimental setup, see Supplementary Information). For this setup we have measured switching times of nano-seconds, which we think were limited by our measurement apparatus \cite{13Alonso}. This is much faster than the ion oscillation period, meaning that we can induce quasi-instantaneous changes to the oscillator states (Fig.\ \ref{fig:expseq}a). Using these techniques, we map out the ion-light coupling strength for high excitations, and track the oscillation trajectory of a single ion throughout multiple oscillation cycles.

In order to examine the timing quality and reproducibility of our control, we prepare a scenario which is highly familiar in classical physics. Starting with a particle at rest at the bottom of a harmonic potential, we suddenly displace the latter by a distance $x_\text{d}$. For an instant, the particle finds itself at rest but in an excited state, and subsequently oscillates in the new potential. In our experiments, this situation is reproduced in the quantum regime. In the ideal case, the initial state is the quantum ground state $\ket{0}$, which has a minimum-uncertainty wave-packet of root-mean-squared extension $x_0$. The action of the sudden displacement produces a coherent state $\ket{-\alpha_0}$ with $\alpha_0=x_\text{d}/(2x_0)$, which oscillates at a frequency $\omega_\text{m}$ while maintaining the form of the wave-packet. Switching the potential minimum back to $x=0$ after a time $\Delta t$ leaves the oscillator in a coherent state of size
\be
\abs{\alpha}=\abs{\alpha_0}\sqrt{2\sbra{1-\cos(\omega_\text{m}\Delta t)}}\label{eq:alpha}
\ee
in the original potential (Fig.\ \ref{fig:expseq}b).

The trapped-ion oscillator we consider also has internal degrees of freedom, from which we can isolate a two-state pseudo-spin system ($\ket{\downarrow},\ket{\uparrow}$) with transition frequency $\omega_0$. This transition can be parametrically coupled to the oscillator using electromagnetic radiation at frequency $\omega_\text{L}= \omega_0 + s \omega_\text{m}$. For integer $s$, resonant transitions can be induced on the $s^\text{th}$ sideband between the states $\ket{\downarrow}\ket{n}\leftrightarrow \ket{\uparrow}\ket{n + s}$ with a Rabi frequency \cite{79Wineland}
\be
\Omega_{n,n+s}= \Omega_0\abs{\bra{n+s}\Exp{\I k_x \hat{x}}\ket{n}},\label{eq:Rabisimp}
\ee
where $\Omega_0$ is a constant which contains the internal-state coupling strength and the electric field amplitude, and $k_x$ is the size of the radiation wavevector projected on the oscillation direction. For $k_x x_0 \ll 1$ and at low excitations, a small parameter expansion of the exponential is sufficient to describe the resulting dynamics. This is the regime in which most quantum control experiments with trapped ions have been operated in the past, and as a result coherent control has only been demonstrated for $\abs{s} \leq 2$. In the work presented below, we map out the excitations for $0\leq s\leq 5$, confirming the dependence of the light-atom interaction strength for excitations up to $n \approx$~10,000.

\ifthenelse{\boolean{isnature}}{}{\capfigii}

In our experiments, the oscillator is the axial motion of a single \caf ion confined in a radio-frequency trap with a frequency between $\omega_\text{m}/(2\pi)\approx \unit{2.35}{MHz}$ and $\unit{2.53}{MHz}$, and corresponding values of $x_0 \sim 7$~nm. We use the dipole-forbidden transition at wavelength $\lambda\approx\unit{729}{nm}$ to define a pseudo-spin system between levels $\ket{\downarrow}\equiv\ket{L=0,J=1/2,M_J=-1/2}$ and $\ket{\uparrow}\equiv\ket{L=2,J=5/2,M_J=-5/2}$. The wavevector of the laser addressing this transition makes an angle $\theta\approx\unit{45}{deg}$ with the motional direction, resulting in a Lamb-Dicke parameter $\eta = k_x x_0 \sim 0.044$. Each experimental sequence begins by initializing the trapped-ion oscillator close to its ground state using a combination of Doppler and electromagnetically-induced-transparency cooling \cite{00Roos}. We measure a typical mean excitation after cooling to be $\nbar_\text{th}= 0.20(6)$. Error bars throughout this paper are given as standard error of the mean.

In a first set of experiments we characterized our control using two successive displacements of the potential well (Fig.\ \ref{fig:expseq}c). At $t = 0$, we switch the center of the potential from 0 to $x_\text{d}$, resulting in the ion being excited and subsequently oscillating in the displaced potential well. At $t = \Delta t$ we reverse this change, bringing the potential back to its original position. In the ideal case, the ion is then in a coherent state of size $|\alpha|$ given by equation (\ref{eq:alpha}). In order to read out the oscillator state, we then perform an analysis using an optical probe pulse of duration $t_\text{p}$ which is resonant with the $s^\text{th}$ sideband. This is followed by a detection of the ion's internal state using resonance fluorescence (Supplementary Information). By repeating the experiment a large number of times for each value of $t_\text{p}$, we obtain an estimate of the probability of finding the ion spin down as a function of time. This agrees well with the functional form
\be
P_\downarrow(t_\text{p})=\frac{1}{2}\sum_{n\geq0} p_{\nbar_\text{th},\abs{\alpha}}(n)\sbra{1+\Exp{-\Gamma t_\text{p}}\cos\rbra{\Omega_{n,n+s}t_\text{p}}},
\label{eq:Pdispth}
\ee
where $p_{\nbar_\text{th},\abs{\alpha}}(n)$ is the occupation of the $n^\text{th}$ number state for a displaced thermal state with thermal mean quantum number $\nbar_\text{th}$ and displacement $|\alpha|$ (this distribution is given in the Supplementary Information). The effects of spin and motional decoherence during the probe pulse are accounted for by an exponential decay at rate $\Gamma$ \cite{96Meekhof}. Experiments were performed for a range of values of $\Delta t$, in each case taking data using the first sideband $s = 1$ for $0 < t_\text{p} < \unit{100}{\micro s}$. For a single value of $\Delta t$, the data do not allow the extraction of $\abs{\alpha}$, $\Omega_0$ and the decay parameter $\Gamma$ independently. We therefore probe first the non-displaced state directly after cooling. Fitting using equation (\ref{eq:Pdispth}) with $\abs{\alpha}=0$ and $\nbar_\text{th}=0.2$, we determine $\Omega_0/(2\pi)=\unit{181(1)}{kHz}$ and $\Gamma= \unit{2.2(3)}{ms^{-1}}$. These are then fixed in fits to the data for each value of $\Delta t$, from which we extract the corresponding displacement $|\alpha|$. The resulting values are plotted in Fig.\ \ref{fig:lobes}a. Also shown is a fitted curve using equation (\ref{eq:alpha}), which
allows us to determine the oscillation frequency of the ion $\omega_\text{m}/(2\pi)=\unit{2.3505(6)}{MHz}$ and amplitude $\alpha_0= 5.11(1)$, corresponding to $x_\text{d}=\unit{75.0(1)}{nm} $ (more details of this analysis can be found in the Supplementary Information). We observe in the results in Fig.\ \ref{fig:lobes}a that the amplitude of each oscillation of the ion appears to be different. We attribute this to drifts in the laser intensity at the ion, which result in shifts in the deduced value of $\abs{\alpha}$ (this drift is illustrated by the bounding curves in Fig.\ \ref{fig:lobes}a, see Supplementary Information for further details).

In order to verify that the switching between potentials does not produce additional effects beyond a simple displacement, we examine our ability to catch the ion in the ground state after a single cycle of oscillation. In an independent measurement we examined data for the region around $\Delta t = 2 \pi/\omega_\text{m}$ for $\omega_\text{m}/(2\pi)\approx\unit{2.53}{MHz}$ and an intermediate coherent state of $\alpha_0 \approx 85$. To faithfully return these states to the origin after a single trap cycle we need to trigger the experimental sequence from the phase of the radio-frequency trap drive. This can be attributed to a pseudopotential gradient along the axis of our trap \cite{03Leibfried2}. The results are presented in Fig.\ \ref{fig:lobes}b and show $|\alpha| = 0.05(10)$ for $\Delta t \approx 2 \pi/\omega_\text{m}$. Figure \ref{fig:lobes}c provides a direct comparison between data for a non-displaced ion with $\nbar_\text{th}=0.15(3)$, and one with the same starting temperature which has been displaced to $\alpha_0 \approx 85$ and returned one cycle later. A fit to the data for the latter gives $\nbar_\text{th}=0.21(3)$ and consistent values for $\Omega_0$ and $\Gamma$.




\ifthenelse{\boolean{isnature}}{}{\capfigiii}

Having characterized our control, we perform a second set of experiments to map out the interaction between the ion and the light field as a function of the motional excitation. In contrast to the work described above, we send the probe pulse to the ion while the potential is centered at $x_\text{d}$. The potential well is displaced back to $x = 0$ after the coherent probe pulse has been applied, timing this second displacement to happen once the ion has completed an integer number of oscillation periods. In this way, we prevent the high oscillator excitation from affecting the internal state detection. Experimental measurements of $P_\downarrow$ are shown in Fig.\ \ref{fig:2Dplots}a for sidebands between $s = 0$ and $s = 5$ and displacements up to $x_\text{d} \approx\unit{1.5}{\micro m}$, which corresponds to $\alpha_0 \approx 100$ and $\nbar \approx 10,000$ at the trap frequency of \unit{2.35}{MHz}. For each value of $x_\text{d}$ we probe $P_\downarrow$ using laser-pulse durations between $1.4$ and $\unit{80}{\micro s}$. To analyze these measurements, we start by determining $\Omega_0/(2\pi)=\unit{204.3(1)}{kHz}$ and $\nbar=0.21(8)$ from measurements with the non-displaced state using the $s=0$ and $s=1$ sidebands respectively. We then calibrate the displacement sizes using the data taken for the carrier transition $s = 0$, fixing $\Gamma=0$  and floating only $\abs{\alpha}$ in equation (\ref{eq:Pdispth}) (for further details, see Supplementary Information). These calibrated values were then fixed, and data for each of the other sidebands was fitted as a full set using a model including motional-excitation-dependent AC Stark shifts due to nearby atomic transitions and a systematic detuning to account for miscalibration (Supplementary Information). We find that the AC Stark shifts become relevant for $s = 4$ and $s = 5$ at high motional excitations, because the frequencies of these transitions are close to the seventh and sixth red motional sidebands respectively of an atomic transition which is \unit{25.74}{MHz} away from the driven transition. Two-dimensional plots showing the fit results are shown in Fig.\ \ref{fig:2Dplots}b below the respective dataset.

In order to verify the predicted form of the matrix elements in equation (\ref{eq:Rabisimp}), we take a Fourier transform of $P_\downarrow(t_\text{p})$ for each value of $x_\text{d}$. This gives us the Rabi frequency distribution, which is peaked due to the limited range of number states which contribute significantly to the coherent state (Supplementary Information). Fig.\ \ref{fig:2Dplots}c shows the mean Rabi frequency obtained from each Fourier transform versus the amplitude of the state. We overlay this with theoretical curves using equation (\ref{eq:Rabisimp}) which include modifications for the offset detunings and AC Stark shifts, showing good agreement between experiment and theory over the full range of motional occupations. The minima in Rabi frequencies observed for each set of sideband data are separated by between 515 and \unit{525}{nm}. These arise due to the large amplitude of the ion's motion, which spatially samples multiple wavelengths of the light. The separation between minima is expected to approach the effective wavelength of the light as projected on the trap axis for $x_\text{d} \gg \lambda$. In our case this value is  $\lambda/(2 \cos\theta) \approx \unit{515}{nm}$ (for more details see Supplementary Information). We note that though the $s = 5$ sideband would take 2 minutes to invert the spin if driven in the ground state, the modulation of the light due to the large ion oscillation allows us to do this in less than \unit{10}{\micro s}.


The work presented has a number of possible applications for quantum information science. It may be used to speed-up the transport of trapped-ion qubits in scalable quantum information processing \cite{02Kielpinski, 12Bowler, 12Walther}, where the transport would proceed by switching the position of the potential well to two positions $x_\text{d}$ and $2 x_\text{d}$ at times which are separated by $\Delta t = \pi/\omega_\text{m}$ \cite{13Alonso}. This would result in the ion being ``caught'' in the ground state of the potential centered at $2 x_\text{d}$, with the transport taking half a period of the ion's oscillation. We expect that large squeezed states could be achieved through sudden trap frequency changes, in this case separated by delays of one quarter of the trap period \cite{13Alonso}. Bang-bang control routines based on sudden changes to the trap frequency have been proposed for suppressing motional decoherence effects \cite{01Vitali}. Using ion chains, the ability to change the trapping potential on timescales which are fast compared to the characteristic frequencies of the interaction between ions has been proposed as a method for generating high levels of continuous variable entanglement by means of the Coulomb interaction \cite{01Gottesman,09Serafini,09Serafini2}.

\ifthenelse{\boolean{isnature}}{
	\clearpage
	\newpage
}

\ifthenelse{\boolean{isnature}}{
\clearpage
\newpage	

\capfigi

\capfigii

\capfigiii

}{}

\ifthenelse{\boolean{isnature}}{
\clearpage
\newpage
	
\begin{addendum}
 \item [Acknowledgements] We thank Daniel Kienzler and Simon Ragg for comments to the manuscript, and Ludwig de Clercq, Frieder Lindenfelser, Daniel Kienzler and Roland Habl\"utzel for contributions to the experimental apparatus. JA thanks Peter Leek for guidance during first steps of trap fabrication. We acknowledge support from the Swiss National Science Foundation under grant numbers 200021\_134776, 200020\_153430 and COST-C12.0118, and from ETH Research Grant ETH-18 12-2.
 \item[Author Contributions] Experimental data were taken by JA and FML, using an apparatus primarily built up by FML and JA, and with significant contributions from ZUS, MF, MM, BCK and VN. Data analysis was performed by FML and JA. The paper was written by JA and JPH, with input from all authors. The work was conceived by JPH and JA.
 \item[Competing Interests] The authors declare that they have no competing financial interests.
 \item[Correspondence] Correspondence and requests for materials should be addressed to \\
 J. Alonso (email: alonso@phys.ethz.ch) or J. P. Home (email: jhome@phys.ethz.ch).
\end{addendum}
}
{
\textbf{Acknowledgement:}  We thank Daniel Kienzler and Simon Ragg for comments to the manuscript, and Ludwig de Clercq, Frieder Lindenfelser, Daniel Kienzler and Roland Habl\"utzel for contributions to the experimental apparatus. JA thanks Peter Leek for guidance during first steps of trap fabrication. We acknowledge support from the Swiss National Science Foundation under grant numbers 200021\_134776, 200020\_153430 and COST-C12.0118, and from ETH Research Grant ETH-18 12-2.\\

\textbf{Author Contributions:} Experimental data were taken by JA and FML, using an apparatus primarily built up by FML and JA, and with significant contributions from ZUS, MF, MM, BCK and VN. Data analysis was performed by FML and JA. The paper was written by JA and JPH, with input from all authors. The work was conceived by JPH and JA.\\

The authors declare that they have no competing financial interests.
}

\clearpage
\newpage

\newcommand{\capfigiv}{
	\ifthenelse{\boolean{isnature}}{\begin{figure*}}{\begin{figure}}
		\includegraphics[width = .95\columnwidth]{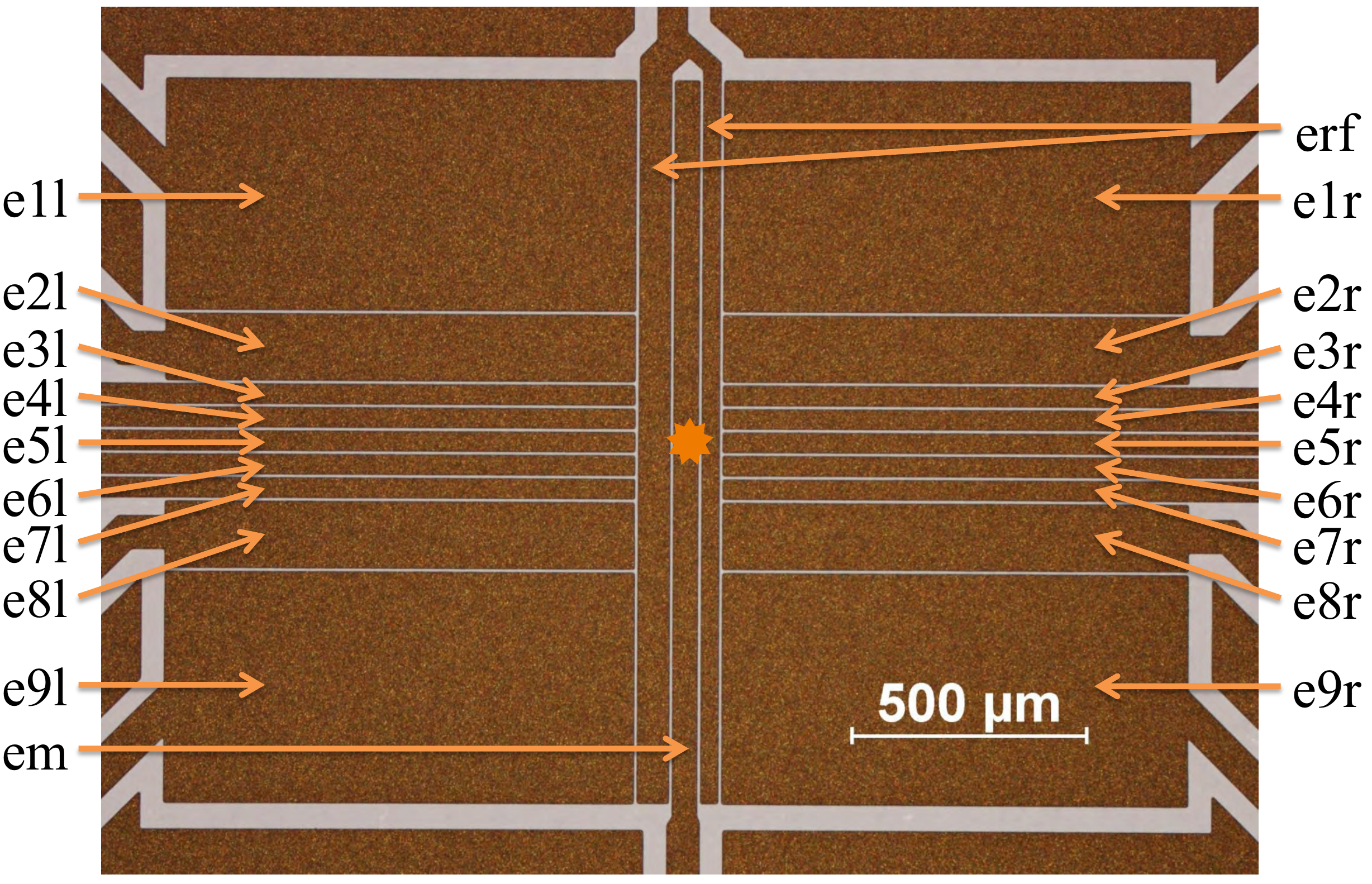}
		\caption{\textbf{Surface-electrode trap:} Linear Paul trap in final fabrication steps. Ions are trapped in the position denoted with a star, approximately \unit{50}{\micro m} from the trap surface.}
		\label{fig:trap}
	\ifthenelse{\boolean{isnature}}{\end{figure*}}{\end{figure}}
}

\newcommand{\capfigv}{
	\begin{figure*}
		\includegraphics[width = 1.95\columnwidth]{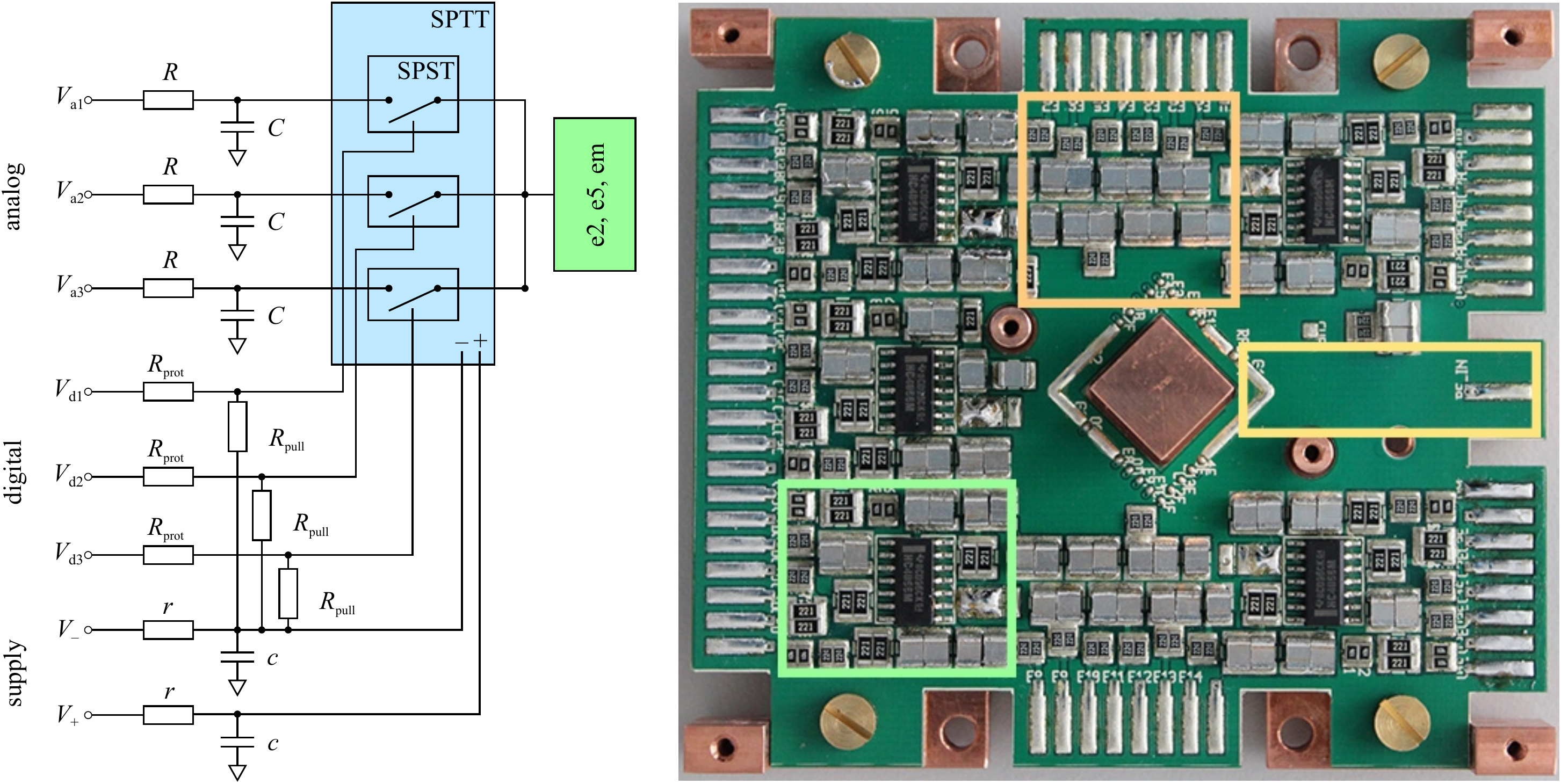}
		\caption{\textbf{Cryogenic electronics:} a) Realization of a SPTT switch for a switchable control electrode.  $r$ and $c$ form low-pass filters for the supply lines of the CMOS switches ($V_+$, $V_-$), with a cutoff frequency of $\approx\unit{7}{kHz}$. $R$ and $C$ form low-pass filters for the input DC voltages ($V_\text{a}$), with a cutoff frequency of $\approx\unit{7}{Hz}$. As usual in rf traps, the capacitors have the additional function of rf-grounding the control electrodes. $R_\text{prot}$ and $R_\text{pull}$ are over-current protection and pull-down resistors for the digital control lines of the SPSTs. $V_\text{a}$ is passed on to the electrode by closing the SPST with a digital high/low level through the control line $V_\text{d}$. b) Picture of the CEB. The circuitry relative to one of five switches is highlighted in green, a group of seven RC filters in orange, and the position of the rf track (inside the PCB) in yellow. The CEB is mounted on a copper structure for thermal contact to the cold plate.}
		\label{fig:CEB}
	\end{figure*}
}

\newcommand{\capfigvi}{
	\ifthenelse{\boolean{isnature}}{\begin{figure*}}{\begin{figure}}
		\includegraphics[width = .95\columnwidth]{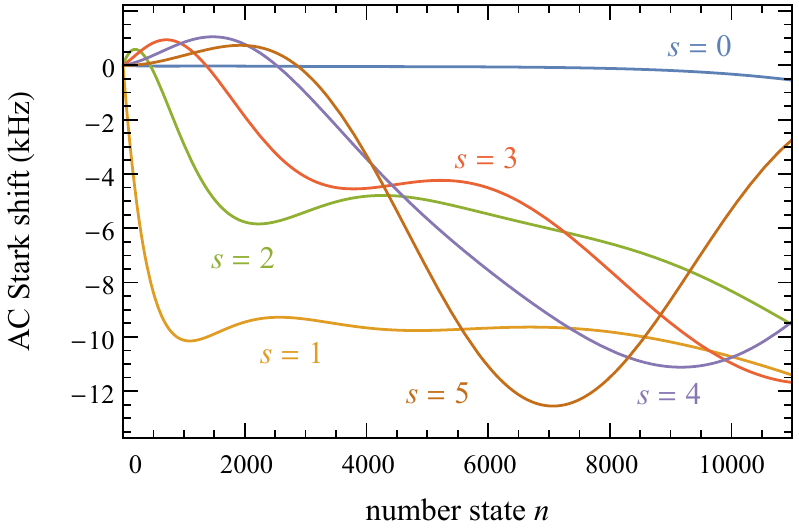}
		\caption{\textbf{AC Stark shifts:} Energy level corrections due to off-resonant coupling to motional sidebands other than the one driven, as well as to the carrier and sidebands of the $\ket{\downarrow}\leftrightarrow\ket{L=2,J=5/2,M_J=3/2}$ transition at our $\unit{3.83}{G}$ magnetic field when $\delta=0$ at $n=0$.}
		\label{fig:acstark}
	\ifthenelse{\boolean{isnature}}{\end{figure*}}{\end{figure}}
}

\newcommand{\capfigvii}{
	\ifthenelse{\boolean{isnature}}{\begin{figure*}}{\begin{figure}}
		\includegraphics[width = .95\columnwidth]{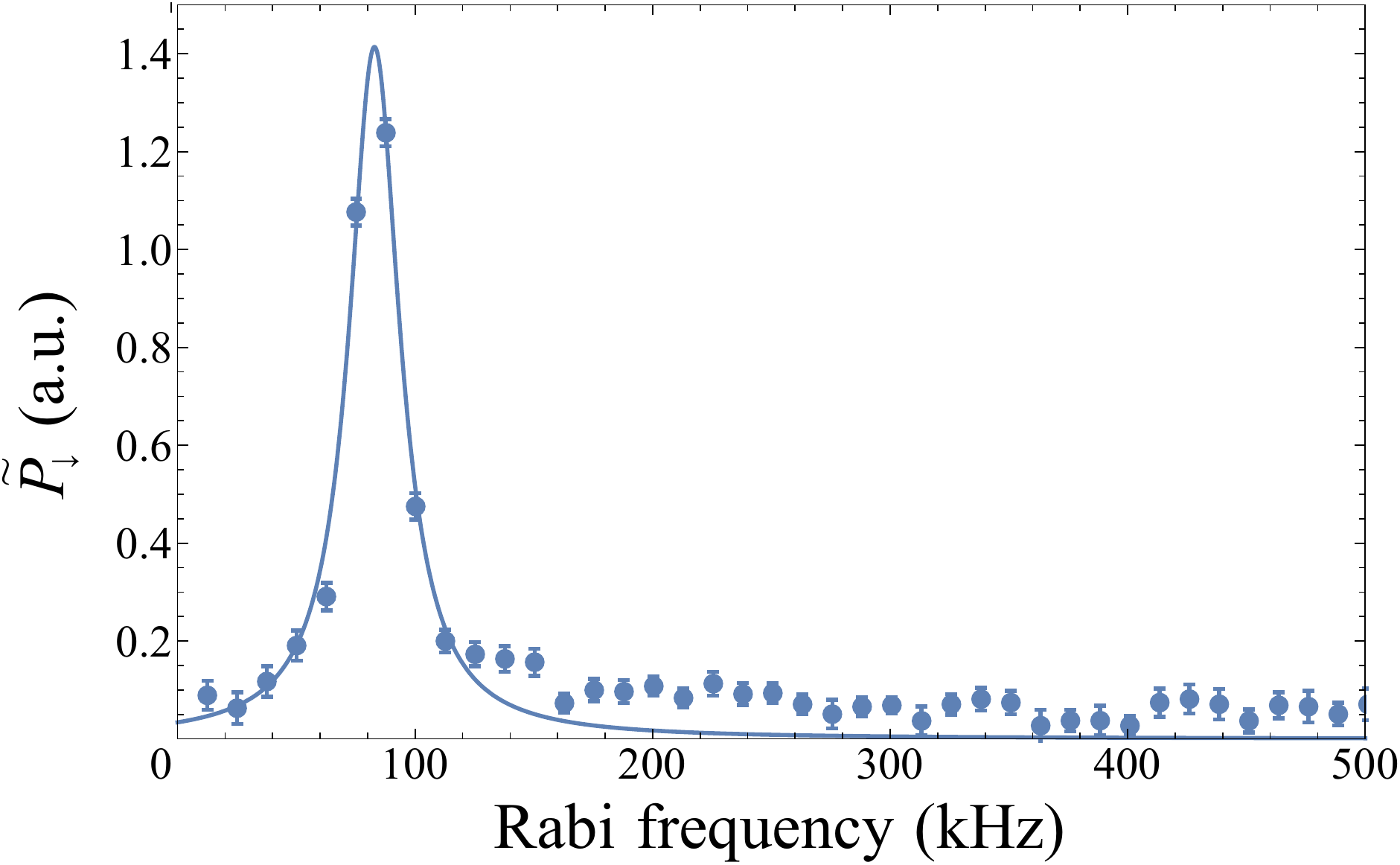}
		\caption{\textbf{Rabi-frequency spectrum:} The data points and curve are obtained as explained in the text. This plot is for the carrier transition ($s=0$) and a displacement $x_\text{d}\approx\unit{660}{nm}$ ($\alpha_0\approx44$).}
		\label{fig:PdownOmega}
	\ifthenelse{\boolean{isnature}}{\end{figure*}}{\end{figure}}
}

\newcommand{\captabi}{
	\begin{table}
		\caption{Summary of the measured performance of the switching-electronics at 300 and \unit{4}{K}. Off and on resistance are measured directly at the switch. The power consumption was measured at a switching rate of \unit{1}{MHz}. The crosstalk indicated is the fraction of power from the digital side leaking to the analog output, and is measured around \unit{1.5}{MHz}. The rise/fall times are given for the 10\%-90\% transition of a \unit{9}{V} step.}
		\label{tab:switches}       
		\begin{center}
			\begin{tabular}{c | c | c}
				& \unit{300}{K} & \unit{4}{K} \\
				\hline
				Off resistance (M$\Omega$) & $>10$ & $>10$ \\
				On resistance ($\Omega$) & 12 & 3.5 \\
				Power consumption (mW) & 8 & 10 \\
				Crosstalk (dB) & -50 & -50 \\
				Rise/Fall time (ns) & 13 & $<5$ \\
			\end{tabular}
		\end{center}
	\end{table}
}

\newcommand{\captabii}{
	\begin{table}
		\caption{Fit results for the light-matter-interaction experiment. The probe-pulse frequencies for sidebands $s=0$ to 2 were pre-calibrated with non-displaced states. Higher-order sidebands are too weak for a direct measurement with non-displaced states, so probe-pulse frequencies for $s>2$ where estimated from the lower sideband frequencies. This explains the large systematic shifts in $\delta_\text{off}$ resulting from the fits.}
		\label{tab:fits2}       
		\begin{center}
			\begin{tabular}{c | c | c }
				$s$ & $\Omega_0/(2\pi)$ (kHz) & $\delta_\text{off}/(2\pi)$ (kHz) \\
				\hline
				0 & 205 & 0 \\
				1 & 211 & 4 \\
				2 & 218 & -4 \\
				3 & 224 & -12 \\
				4 & 226 & -23 \\
				5 & 226 & -32 \\
			\end{tabular}
		\end{center}
	\end{table}
}

\newcommand{\sectitle}{Supplementary material}
\ifthenelse{\boolean{isnature}}{\section*{\sectitle}}{\section{\sectitle}}

\newcommand{\subsectitle}{Experimental setup}	
\ifthenelse{\boolean{isnature}}{\textbf{\subsectitle.}}{\subsection{\subsectitle}}
\ifthenelse{\boolean{isnature}}{}{\capfigiv}
Our experiments are performed using a surface-electrode linear radio-frequency (rf) trap in a 5-wire asymmetric configuration \cite{05Chiaverini2,11Amini} (Fig.\ \ref{fig:trap}). The pseudo-potential null line is $\approx\unit{50}{\micro m}$ above the trap chip. The main trapping zone, denoted with a star in the figure, is between the pair of electrodes e5, closer to e5r than e5l due to the asymmetry in the width of the rf electrodes. Electrodes e2, e8 and em are each connected to the output of a Single-Pole-Triple-Throw (SPTT) switch mounted on the Cryo-Electronics Board (CEB, Fig.\ \ref{fig:CEB}b) and within \unit{3}{cm} of the trap chip. Voltage offsets of less than a volt at these five electrodes suffice to carry out our experiments. The trap is driven with an rf amplitude of $\approx\unit{100}{V}$ and a frequency of $\approx\unit{93}{MHz}$, leading to radial secular motion at $\approx 4$ and $\approx\unit{7}{MHz}$ at an axial frequency of $\omega_\text{m}/(2\pi)\approx\unit{2.5}{MHz}$.

The trap is placed in a sealed chamber cooled down to $\approx\unit{4}{K}$. Cryogenic setups yield improved vacuum over room-temperature experiments \cite{90Gabrielse}, resulting in longer ion lifetimes. We take advantage of the low outgassing at cryogenic temperatures to use standard Printed-Circuit-Board (PCB) assemblies for in-vacuum electronics. This reduces the technical effort compared to preparing ultra-high-vacuum-compatible electronics for room temperature operation \cite{14Guise}.

\ifthenelse{\boolean{isnature}}{}{\capfigv}

The CEB holds the fast switching electronics, 30 low-pass filters for the analog input lines, and a track for guiding the trap rf drive from a quarter-wave helical resonator to the trap chip. For voltage switching we use a commercial CMOS integrated circuit (CD74HC4066M, from Texas Instruments), which implements four bilateral Single-Pole-Single-Throw (SPST) switches. The additional circuitry to implement a SPTT switch is shown in Fig.\ \ref{fig:CEB}a. The CEB includes five such copies, one per switchable electrode. Measured parameters of this circuit are shown in Tab.\ \ref{tab:switches}. Resistors and capacitors are attached in parallel pairs in order to have a spare connection in the event of solder-point damage during cool-down. To ensure cryo-compatibility, all resistors are thin-film and capacitors are from the Panasonic ECHU(X) series. Otherwise, the PCB design and soldering techniques do not differ from those used to prepare standard non-vacuum boards.

\ifthenelse{\boolean{isnature}}{}{\captabi}

The digital pulses for controlling the switches are produced by room-temperature electronics based on a multichannel delay/pulse generator (P400 from Highland Technology). Sequences of pulses are triggered by a single TTL line locked to the phase of the rf drive and generated by a Field-Programmable-Gate-Array (FPGA), which we also use for the generation of laser-pulse sequences.

\renewcommand{\subsectitle}{Cooling and Detection}	
\ifthenelse{\boolean{isnature}}{\textbf{\subsectitle.}}{\subsection{\subsectitle}}
The experimental sequences are depicted in Fig.\ \ref{fig:expseq}c-d. Ground-state cooling is achieved in two stages. The first involves  Doppler cooling into the Lamb-Dicke regime by applying a laser $10$~MHz detuned below the $S_{1/2}\leftrightarrow P_{1/2}$ transition at \unit{397}{nm} for \unit{500}{\micro s}. We then follow this with electromagnetically-induced-transparency cooling, involving the application of two \unit{397}{nm} laser fields polarized so that they drive simultaneously the $\sigma^-$ and $\pi$ transitions between the $S_{1/2}$ and $P_{1/2}$ Zeeman sub-levels. With suitable combinations of laser powers and frequency detunings, this allows us to cool the ion close to the ground state of motion \cite{00Roos}. In our setup we reach a steady state at $\nbar_\text{th}\approx 0.2$ after \unit{100}{\micro s}, limited by the heating rate of the ion, which is $1-2$~quanta$/$ms from the quantum ground state. After cooling, the state is initialized to $\ket{\downarrow}$ by optically pumping with a $\sigma^-$-polarized resonant \unit{397}{nm} beam. In all these pulses, \unit{866}{nm} light is used to repump the population from $D_{3/2}$ back to $P_{1/2}$. Coherent operations between the $S_{1/2}$ and $D_{5/2}$ states are performed with a narrow-linewidth \unit{729}{nm} laser. These couple the motional state to the spin system, which is subsequently read out by state-dependent fluorescence with resonant \unit{397}{nm} and \unit{866}{nm} light. In a detection time interval of \unit{500}{\micro s}, we detect a mean number of 2 photons for an ion initially prepared in $\ket{\uparrow}$, and 25 photons if the ion is prepared in $\ket{\downarrow}$.

\renewcommand{\subsectitle}{Displaced thermal states}	
\ifthenelse{\boolean{isnature}}{\textbf{\subsectitle.}}{\subsection{\subsectitle}}
For a thermal state with average phonon number $\nbar_\text{th}$ displaced by $\alpha$, the occupation of the $n^\text{th}$ number state is \cite{73Boiteux}
\be
\notag p_{\nbar_\text{th},\abs{\alpha}}(n)&=&\Exp{-\abs{\alpha}^2}\sum_{m\geq0} \frac{\nbar_\text{th}^n}{(\nbar_\text{th}+1)^{n+1}}n!m!\abs{\alpha}^{2(n-m)}\\
&& \times \abs{\sum_{j=I}^m\frac{(-1)^j\abs{\alpha}^{2j}}{j!(m-j)!(n-m+j)!}}^2,\label{eq:palpha}
\ee
with $I=\max\cbra{0,m-n}$.

\renewcommand{\subsectitle}{Light-atom interaction}	
\ifthenelse{\boolean{isnature}}{\textbf{\subsectitle.}}{\subsection{\subsectitle}}
The coupling between the internal electronic state and the ion's motion is described by the Hamiltonian \cite{03Leibfried2}
\be
\op{H}_\text{c}^\text{I}=\frac{\hbar\Omega_0}{2}\splus\Exp{\I\eta(\destroy\Exp{-\I\omega_\text{m} t_\text{p}}+\create\Exp{\I\omega_\text{m} t_\text{p}})}\Exp{-\I\delta t_\text{p}}+\text{H.c.}\label{eq:Ham}
\ee
Here, $\splus=\ket{\uparrow}\bra{\downarrow}$ is the spin-flip operator producing transitions from $\ket{\downarrow}$ to $\ket{\uparrow}$ at the Rabi frequency $\Omega_0/(2\pi)$, the Lamb-Dicke parameter $\eta=k_x x_0$ relates the motional ground-state wave-packet size $x_0=\sqrt{\hbar/(2M\omega_\text{m})}$ to the projection of the radiation wave vector $\vec{k}$ on the trap axis, $M$ is the ion's mass, and $\delta$ is the laser detuning from the carrier transition at frequency $\omega_0$. For integer values of $s$, laser light detuned by $\delta\approx s\omega_\text{m}$ drives near-resonant transitions on the $s^\text{th}$ motional sideband $\ket{\downarrow}\ket{n}\leftrightarrow \ket{\uparrow}\ket{n + s}$. The corresponding matrix element is proportional to $\abs{\bra{n+s}\Exp{\I k_x \op{x}}\ket{n}}$ and its associated Rabi frequency can be evaluated analytically \cite{79Wineland}:
\be
\notag \Omega_{n,n+s} &=&\Omega_0\abs{\bra{n+s}\Exp{\I\eta(\create+\destroy)}\ket{n}}\\
&=& \Omega_0\Exp{-\eta^2/2}\eta^{|s|}\sqrt{\frac{n_<!}{n_>!}}\abs{L_{n_<}^{|s|}(\eta^2)},\label{eq:Rabi}
\ee
where, $n_<$ ($n_>$) stands for the lesser (greater) of $n+s$ and $n$, and $L_{n}^{\alpha}(x)$ are generalized Laguerre polynomials.

\ifthenelse{\boolean{isnature}}{}{\capfigvi}

\renewcommand{\subsectitle}{Dependence of deduced oscillation amplitude on carrier Rabi frequencies}	
\ifthenelse{\boolean{isnature}}{\textbf{\subsectitle.}}{\subsection{\subsectitle}}
For the first set of experiments we carried out the sequence in Fig.\ \ref{fig:expseq}c and obtained the results in Fig.\ \ref{fig:lobes}. Here we focus on one of the consequences of the slow drifts in laser power, which are not included in the model used for theoretical calculations.

From a first fit of a non-displaced state we obtain $\Omega_0/(2\pi)=\unit{181(1)}{kHz}$ and $\Gamma=\unit{2.2(3)}{ms^{-1}}$, using equation (\ref{eq:Pdispth}) with $\abs{\alpha}=0$ and $\nbar_\text{th}=0.2$. From subsequent fits for each value of $\Delta t$ we obtain the coherent-state sizes plotted in Fig.\ \ref{fig:lobes}a. Fitting these to equation (\ref{eq:alpha}), we determine $\alpha_0=5.11(1)$ and $\omega_\text{m}/(2\pi)=\unit{2.3505(6)}{MHz}$. However, we note that over the time required to take this data the Rabi frequency varies over a range $\delta\Omega_0/(2\pi)\approx\pm\unit{3.2}{kHz}$. The value of $\abs{\alpha}$ determined from fits is strongly correlated to the Rabi frequency used, which accounts for the fluctuations observed in Fig.\ \ref{fig:lobes}a. To illustrate this we repeat the fits fixing the Rabi frequency to the limiting values $\Omega_0\pm\delta\Omega_0$ and floating $\alpha_0$ and $\omega_\text{m}$. This yields $\alpha_0=5.00(1)$ for $\Omega_0+\delta\Omega_0$, and $\alpha_0=5.22(1)$ for $\Omega_0-\delta\Omega_0$, with $\omega_\text{m}/(2\pi)=\unit{2.3505(6)}{MHz}$ in both cases. The boundaries of the shaded area in Fig.\ \ref{fig:lobes}a are obtained by inserting these numbers into equation (\ref{eq:alpha}).

\renewcommand{\subsectitle}{Motional-state dependent AC Stark shifts}	
\ifthenelse{\boolean{isnature}}{\textbf{\subsectitle.}}{\subsection{\subsectitle}}
The model used for calculating the expected $P_\downarrow$ in our experiment (Fig.\ \ref{fig:2Dplots}) includes a detuning from the driven motional sideband as well as an AC Stark shift due to off-resonant coupling to transitions other than the one probed. The latter is given by
\be
\notag \delta_\text{AC}(n,s)&=&\sum_{s'\neq s}\frac{\Omega_{n,n+s}^2}{2\omega_\text{m}(s'-s)}\\
&& + \sum_{s'}\frac{(\Omega_{n,n+s}/W_{-\frac{1}{2}\rightarrow\frac{3}{2}})^2}{4(\Delta_{-\frac{1}{2}\rightarrow\frac{3}{2}}+\omega_\text{m}(s'-s))} \label{eq:acstark}
\ee
and plotted in Fig.\ \ref{fig:acstark} for our experimental parameters. The first term arises from coupling to sidebands of the spin transition, while the second is due to the fact that the \unit{729}{nm} laser drives off-resonantly the secondary transition $\ket{\downarrow}\leftrightarrow\ket{L=2,J=5/2,M_J=3/2}$. The coupling strength to the secondary transition is reduced relative to that of the resonantly driven transition by a factor $W_{-\frac{1}{2}\rightarrow\frac{3}{2}}=\sqrt{5}$. At our magnetic field of $\approx\unit{3.83}{G}$, the frequency gap between both carrier transitions is $\Delta_{-\frac{1}{2}\rightarrow\frac{3}{2}}/(2\pi)\approx\unit{25.74}{MHz}$. This is close to an integer times the motional frequency, causing the $s=4$ (5) sideband of the main transition to be almost resonant with the $s=-7$ ($-6$) sideband of the secondary transition, and therefore strongly AC Stark shifting the $\ket{\downarrow}$ state.  For a total detuning $\delta_\text{tot}(n,s)=\delta_\text{off}(s)+\delta_\text{AC}(n,s)$, equation (\ref{eq:Pdispth}) can be re-written as
\ifthenelse{\boolean{isnature}}{}{\begin{widetext}}
	\begin{equation}
	P_\downarrow(\nbar_\text{th},\abs{\alpha},\Omega_0,s,t_\text{p})=1-\sum_{n\geq0} p_{\nbar_\text{th},\abs{\alpha}}(n)\frac{\Omega_{n,n+s}^2}{\Omega_{n,n+s}^2+\delta_\text{tot}^2 (n,s) }\sin^2    \rbra{\sqrt{\Omega_{n,n+s}^2+\delta_\text{tot}^2(n,s)}~\frac{t_\text{p}}{2}},
	\label{eq:Pdispth2}
	\end{equation}
\ifthenelse{\boolean{isnature}}{}{\end{widetext}}
where we have ignored the exponential-decay term due to decoherence, since our probe times go up to \unit{80}{\micro s} and we found previously $\Gamma\approx\unit{2}{ms^{-1}}$.

\renewcommand{\subsectitle}{Light-matter-interaction data analysis}	
\ifthenelse{\boolean{isnature}}{\textbf{\subsectitle.}}{\subsection{\subsectitle}}
\ifthenelse{\boolean{isnature}}{}{\captabii}
For this experiment we carried out the sequence in Fig.\ \ref{fig:expseq}d. The results are shown in Fig.\ \ref{fig:2Dplots}a. To analyze the data and compare it to the theoretical model, we first calibrated the displacements with the data obtained for the carrier ($s=0$). We fitted the measurement results for each value of $x_\text{d}$ to equation (\ref{eq:Pdispth2}) while fixing $\Omega_0$ and $\nbar$ to values determined previously by driving Rabi oscillations on the $s=0$ and $s=1$ transitions, respectively. This yielded a coherent-state size which we converted into displacements according to $\alpha_0=x_\text{d}/(2x_0)$. For the rest of the sidebands we used the calibrated displacements and fitted the measurement data to find $\Omega_0$ and $\delta_\text{off}$ for each individual sideband. The best fit parameters are given in Tab.\ \ref{tab:fits2} and yield the results in Fig.\ \ref{fig:2Dplots}b-c.

To analyze the evolution of the mean Rabi frequency for a given driven sideband $s$ and as a function of the state size (Fig.\ \ref{fig:2Dplots}c), we calculate the discrete Fourier transform of the measured $P_\downarrow(t_\text{p})$ and propagate the shot-noise uncertainties to the frequency domain according to \cite{97Fornies}. The number-state distribution $p_{\nbar_\text{th},\abs{\alpha}}(n)$ for low values of $\nbar_\text{th}$ is narrow compared to the features which result from equation (\ref{eq:Rabi}), so the frequency spectrum shows a single peak at the mean Rabi frequency (Fig.\ \ref{fig:PdownOmega}). We fit these to symmetric Lorentzian functions, from which we determine the center frequencies given in the plots. We compare this to the  theoretical mean Rabi frequency for a displaced thermal state, given by
\be
\bar{\Omega}_\text{th}=\sum_n p_{\nbar_\text{th},\abs{\alpha}}(n) \sqrt{\Omega_{n,n+s}^2+\delta_\text{tot}^2}.\label{eq:meanOmegath}
\ee

\ifthenelse{\boolean{isnature}}{}{\capfigvii}

The separation between minima in mean Rabi frequency depends on the radiation wavelength $\lambda$. For large values of $n$, the Laguerre polynomials in equation (\ref{eq:Rabi}) can be approximated by Bessel functions $J_\alpha$ as \cite{BkSzeg}
\be
L_n^{\abs{s}}(\eta^2)\approx \rbra{\frac{\sqrt{n}}{\eta}}^{\abs{s}} \Exp{\frac{\eta^2}{2}}J_{\abs{s}}(2\eta\sqrt{n}).\label{eq:Bessel}
\ee
The zeros in the right-hand side  correspond to zeros of the Bessel functions, whose separation tends to $\pi$ for large arguments. This implies that the separation between minima in mean Rabi frequencies tends to $\lambda/(2\cos\theta)$ for large values of $x_\text{d}$.

\ifthenelse{\boolean{isnature}}{
	\clearpage
	\newpage
	
	\capfigiv
	
	\capfigv
	
	\capfigvi
	
	\capfigvii
}{}

\ifthenelse{\boolean{isnature}}{
	\clearpage
	\newpage
	
	\captabi
	
	\captabii
}{}

\end{document}